# Fully reversible transition from Wenzel to Cassie-Baxter states on corrugated superhydrophobic surfaces

Robert J. Vrancken[1], Halim Kusumaatmaja[2,3,*], Ko Hermans[1], An M. Prenen[1], Olivier. Pierre-Louis[2,4], Cees W. M. Bastiaansen[1], Dirk J. Broer[1,5]

[1]*Laboratory of Polymer Technology, Eindhoven University of Technology, P.O. Box 513, 5600 MB Eindhoven, The Netherlands*

[2]*The Rudolf Peierls Centre for Theoretical Physics, Oxford University, 1 Keble Road, Oxford OX1 3NP, U.K.*

[3]*Max Planck Institute of Colloids and Interfaces, Science Park Golm, 14424 Potsdam, Germany*

[4]*Laboratoire de Spectrometrie Physique, Universite J. Fourier, Grenoble 1, BP87, 38402 St Martin d'Heres, France.*

[5]*Philips Research Laboratories, High Tech Campus 4, 5656 AE Eindhoven, The Netherlands*

* Corresponding author, e-mail: halim@mpikg.mpg.de

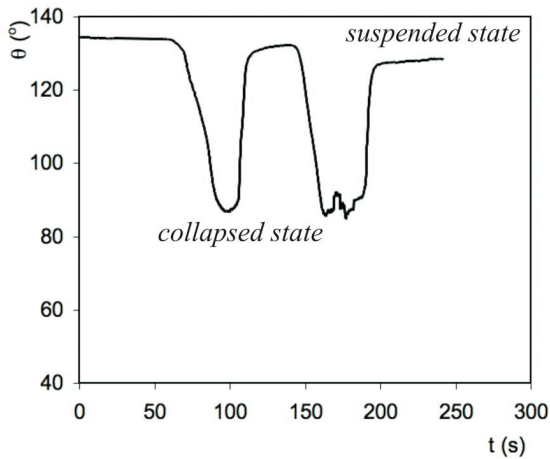
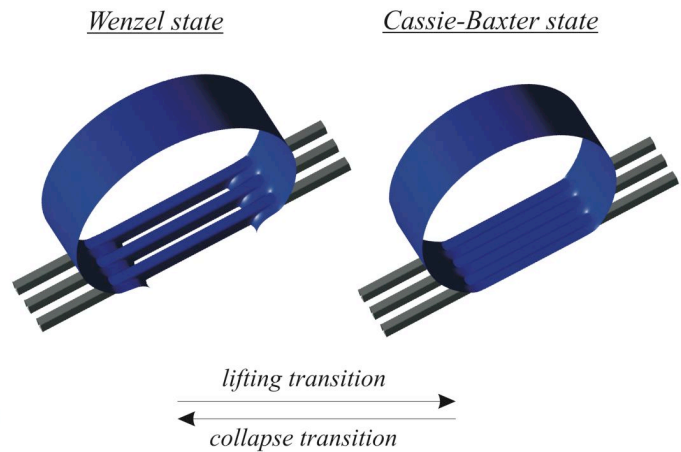

**Table of contents graphic:** *Liquid drops on textured surfaces are extremely mobile when they are supported by composite solid-liquid-air interfaces and immobile when they fully wet the textured surfaces. We demonstrate a new and simple design paradigm, consisting of parallel grooves of appropriate aspect ratio that allows for a controlled, barrierless, and reversible switching of the wetting states upon the application of electrowetting.*

*Abstract*

Liquid drops on textured surfaces show different dynamical behaviors depending on their wetting states. They are extremely mobile when they are supported by composite solid-liquid-air interfaces (Cassie-Baxter state) and immobile when they fully wet the textured surfaces (Wenzel state). By reversibly switching between these two states, it will be possible to achieve large control over the fluid dynamics. Unfortunately, these wetting transitions are usually prevented by surface energy barriers. We demonstrate here a new and simple design paradigm, consisting of parallel grooves of appropriate aspect ratio, that allows for a controlled, barrierless, and reversible switching of the wetting states upon the application of electrowetting. We report a direct observation of the barrierless dynamical pathway for the reversible transitions between the Wenzel (collapsed) and the Cassie-Baxter (suspended) states and present a theory that accounts for these transitions, including detailed lattice-Boltzmann simulations.

*Introduction*

The wetting properties of liquids have been a topic of interest for centuries.[1] Recently, it became clear that both the static and dynamical properties can be controlled via surface patterning. Superhydrophobicity is perhaps the prime example: by making the surface rough, the contact angle of a hydrophobic surface can be increased to close to 180 degrees.[2,3] The two possible states, i.e. collapsed[4] and suspended[5] states, exhibit clear differences in drop mobility. For a variety of applications (e.g. fluid motion control in microfluidics)[6], it is therefore of great interest that reversible transitions between these states can be induced at will. Furthermore, in many cases, it is possible to render the surface effectively hydrophilic when the drop is in the collapsed state and as a result, the contact angle switches from close to 0 (superhydrophilic) to close to 180 degrees (superhydrophobic). This active control of the wetting properties may be induced in a number of ways, for example, by electrowetting, light irradiation, or a change in temperature or pH.[7] Unfortunately, however, a reversible transition from collapsed to suspended states is normally very complicated to achieve, due to the existence of (Gibbs) energy barriers between the states.[8,9] In particular, it is the transition from the collapsed to suspended state that proves problematic, because the base of the liquid drop in the collapsed state cannot detach from the solid surface, as one needs to replace one interface (liquid-solid) with two interfaces (liquid-gas and gas-solid). In other words, this transition must occur via a different pathway. It is therefore our aim in this work to demonstrate a new and simple design paradigm, consisting of parallel grooves of appropriate aspect ratio that allows for a controlled, barrierless, and reversible switching of the wetting states upon the application of electrowetting. We report a direct observation of the barrierless dynamical pathway for the transition from the Wenzel (collapsed) to the Cassie-Baxter (suspended) state and present a theory that accounts for the transition.

So far, to the best of our knowledge, only Krupenkin *et al.* have unambiguously demonstrated a lifting transition by rapid heating of the substrate and evaporating part of the liquid into a vapor blanket beneath the droplet.[10] From here on, we shall term the collapsed to suspended state transition a *lifting* transition and the opposite transition as a *collapse* transition. Although this proves the concept of a lifting transition, the evaporation of liquid is impractical in closed systems such as switchable lenses,[11] microfluidic devices,[12] and displays.[13] Thus, a more practical dynamic pathway is desired for many applications.

Inspired by previous findings that parallel to a corrugated surface the surface patterning does not introduce an energy barrier for the moving liquid-vapor interface,[14,15,16] we investigate the possibility of the lifting transition occurring on such surfaces. To induce the transitions between the aforementioned wetting states, we use electrowetting,[17,18] which has been shown to be applicable on superhydrophobic surfaces.[19,20] We employ lattice Boltzmann simulations to identify the dynamical pathway of the lifting transition and estimate the conditions required to observe it. We then use the outcome of the simulations to guide us in choosing the relevant experimental parameters which allow us to realize reversible and repeatable cycles of lifting and collapse transitions. These transitions are observed indirectly via contact angle measurements and directly by monitoring the motion of liquid/air interfaces under the water droplet during electrowetting.

*Lattice Boltzmann method*

In the simulations reported here, the liquid drop and its surrounding gas are modeled using a one-component, two-phase fluid. The equilibrium properties of the system are described by a Landau free energy functional of the form

$$\Psi = \int_V \left( \psi_b(n) + \frac{\kappa}{2}(\partial_\alpha n)^2 \right) dV + \int_S \psi_s(n_s) dS. \qquad (1)$$

The first term $\psi_b$ describes the bulk free energy of the system, which we choose for convenience to have the form [21]

$$\psi_b(n) = p_c(v_n + 1)^2 (v_n^2 - 2v_n + 3 - 2\beta\tau_w), \qquad (2)$$

where $v_n = (n - n_c)/n_c$, $\tau_w = (T_c - T)/T_c$, and $n$, $n_c$, $T$, $T_c$, and $p_c$ are the local density, critical density, local temperature, critical temperature, and critical pressure of the fluid respectively. This choice of free energy will lead to two coexisting bulk phases of density $n_b = n_c(1 \pm \sqrt{\beta\tau_w})$. $\beta$ is a free parameter in the model that may be used to change the liquid gas density ratio.

The second (gradient) term models the free energy associated with any interfaces in the system. The parameter $\kappa$ is related to two physical quantities:[21] the surface tension via $\gamma_{LG} = (4\sqrt{2\kappa p_c}(\beta\tau_w)^{3/2} n_c)/3$ and the interface width via $\xi = \sqrt{(\kappa n_c^2)/(4\beta\tau_w p_c)}$.

The third (surface) term describes the interactions between the fluid and the solid surface. Following Cahn,[22] the surface energy density is taken to be $\psi_s = -h n_s$, where $n_s$ is the value of the fluid density at the surface. In other words, the strength of interaction is parameterized by the variable *h* and it is related to the contact angle $\theta_e$ by

$$h = 2\beta\tau_w\sqrt{2\kappa p_c}\,\text{sign}\left(\frac{\pi}{2} - \theta_e\right)\sqrt{\cos\left(\frac{\alpha}{3}\right)\left[1 - \cos\left(\frac{\alpha}{3}\right)\right]} \tag{3}$$

where $\alpha = \cos^{-1}(\sin^2\theta_e)$.[21] Minimizing the free energy gives an equilibrium (wetting) boundary condition

$$\kappa\partial_\perp n = -h. \tag{4}$$

Since *h* is related to the contact angle $\theta_e$, the hydrophobicity of the modeled surface can be tuned by varying *h* appropriately.

The hydrodynamics of the drop is described by the continuity and the Navier-Stokes equations

$$\begin{aligned}\partial_t n + \partial_\alpha(nu_\alpha) &= 0, \\ \partial_t(nu_\alpha) + \partial_\beta(nu_\alpha u_\beta) &= -\partial_\beta P_{\alpha\beta} + \nu\partial_\beta\left[n\left(\partial_\beta u_\alpha + \partial_\alpha u_\beta + \delta_{\alpha\beta}\partial_\gamma u_\gamma\right)\right],\end{aligned} \tag{5}$$

where ***u***, ***P***, and $\nu$ are the local velocity, pressure tensor, and kinematic viscosity. The thermodynamics of the drop enters the dynamical equations through the pressure tensor, which can be calculated from the free energy to give

$$\begin{aligned}P_{\alpha\beta} &= \left(p_b(n) - \frac{\kappa}{2}(\partial_\gamma n)^2 - \kappa n\partial_{\gamma\gamma} n\right)\delta_{\alpha\beta} + \kappa(\partial_\alpha n)(\partial_\beta n), \\ p_b(n) &= p_c(\nu_n + 1)^2(3\nu_n^2 - 2\nu_n + 1 - 2\beta\tau_w).\end{aligned} \tag{6}$$

We use the lattice Boltzmann algorithm to solve the continuity and the Navier-Stokes equations.[23] We will not describe the lattice Boltzmann method here and refer interested readers to reference instead.[24,25] In addition to the free energy model we describe here, there are several alternative lattice Boltzmann models.[26,27,28,29] These models have been used extensively to simulate the dynamics of drops on chemically patterned and superhydrophobic surfaces with great success. [8,14,24,25,30,31,32,33,34]

We have chosen the following parameters for our simulations: $\beta$ = 0.1, $\kappa$ = 0.004, $p_c$ = 1/8, $n_c$ = 3.5, and $\tau_w$ = 0.3. These parameters give an interfacial thickness $\xi$ = 1.8, surface tension $\gamma_{LG}$ = 7.7x10$^{-4}$, liquid density $n_L$ = 4.1, and gas density $n_G$ = 2.9 (all in lattice units). We have also used liquid viscosity $\nu$ = 0.69.

It is important to point out that a problem with many mesoscale simulations of liquid–gas systems is that interface widths are too large compared to experiments and the density difference between liquid and gas is too small. The result of this is that time scales are too fast.[34,35] However, this is not critical to the results we present in the paper since we are mainly interested in the criterion for the lifting transition. By changing the values of the interface width, density ratio, surface tension and viscosity, we observe that while the speed of the transition (i.e. how fast it occurs) is clearly affected, the criterion itself (Figure 1d) does not change considerably. Furthermore, previous comparisons between simulations and experiments have shown that the dynamic pathways obtain from the simulations are correct.[14,24,33,34] All these evidence give us confidence that our results in this paper are correct and are not seriously affected by the limitations of the method.

Our lattice Boltzmann simulations are initialized as follows. A hemi-cylindrical drop with radius $R$ = 110 lattice spacings, which is large compared to the size of the grooves, is placed in the Wenzel state on top of a grooved surface for a given contact angle $\theta_e$ (local contact angle without application of electrowetting) and aspect ratio, defined as the height $h$ divided by the width $w$ of the groove. Then, we let the drop equilibrate. For simulation parameters that allow the lifting transition, it is typically achieved in 2 to 4 x 10$^5$ time steps.

Nevertheless, we expect the dynamics of the lifting transition to be slow close to the transition line. To minimize this dynamical effect, we run the lattice Boltzmann simulations for $10^6$ time steps. Simulating a cylindrical drop rather than a full, three dimensional, spherical drop allows us to reduce the system size and hence the computational requirements, while preserving the important physics, in particular a two dimensional curvature of the interface in the direction parallel to the grooves. The simulations are run for different values of $\theta_e$ and $h/w$. The simulation parameters for the surface contact angle and its corrugation are given in Table 1. Further on we also present a comparison to full three-dimensional simulation results.

Table 1: Summary of the lattice Boltzmann simulations parameters.

| $\theta_e$ | w | h | Lifting transition |
|---|---|---|---|
| 110° | 10 | 15 | Yes |
| 110° | 10 | 14 | No |
| 110° | 10 | 12 | No |
| 110° | 10 | 10 | No |
| 110° | 10 | 5 | No |
| 120° | 30 | 30 | Yes |
| 120° | 30 | 27 | Yes |
| 120° | 30 | 24 | Yes |
| 120° | 30 | 22 | No |
| 120° | 30 | 15 | No |
| 120° | 30 | 8 | No |
| 130° | 40 | 20 | Yes |
| 130° | 40 | 16 | No |
| 130° | 40 | 14 | No |
| 130° | 40 | 12 | No |
| 130° | 40 | 10 | No |
| 140° | 50 | 15 | Yes |
| 140° | 50 | 10 | No |
| 140° | 50 | 5 | No |

*Simulation and theoretical results*

In this section, we present the lattice Boltzmann simulation results (Figure 1) and show how this leads us to a simple criterion for the lifting transition to occur. The typical dynamical pathway for the lifting transition shows that, as the drop dewets, the contact line moves considerably slower on top of the barriers than in the grooves, and the shape of the liquid-gas interface that connects the two sides of the groove walls does not change much. As the contact line recedes by *dx*, we are essentially replacing *(2h+w)\*dx* of liquid-solid interface along each groove with *(2h+w)\*dx* of vapor-solid interface and *w\*dx* of liquid-gas interface. Working out the surface energy balance, and employing Young's equation, we find that if:

$$\cos\theta_e < -\frac{1}{1+2(h/w)}, \qquad (7)$$

the transition is energetically favorable. One can immediately see from equation (7) that lifting only happens for $\theta_e > 90°$.

A lifting transition diagram is shown in Figure 1d as a function of the aspect ratio (*h/w*) and contact angle $\theta_e$. Regions where the transition occurs and is suppressed are shown by diamonds and circles respectively. The data points are obtained from the hemi-cylindrical lattice Boltzmann simulations detailed in Table 1, where the solid black line is the analytical result given in equation (7). The agreement between simulation and theory is good, except that in the simulations for a given value of $\theta_e$, the transition occurs at a slightly higher aspect ratio.

This discrepancy can be accounted for by considering the finite size effects of the drop volume. As in the simulations above, we consider a cylindrical drop. Furthermore, for simplicity, we shall make two assumptions. Firstly, the contact angle of the cylindrical cap is close to the Cassie-Baxter contact angle. Secondly, the shape of the liquid-gas interface in the groove can be approximated by a simple plane.

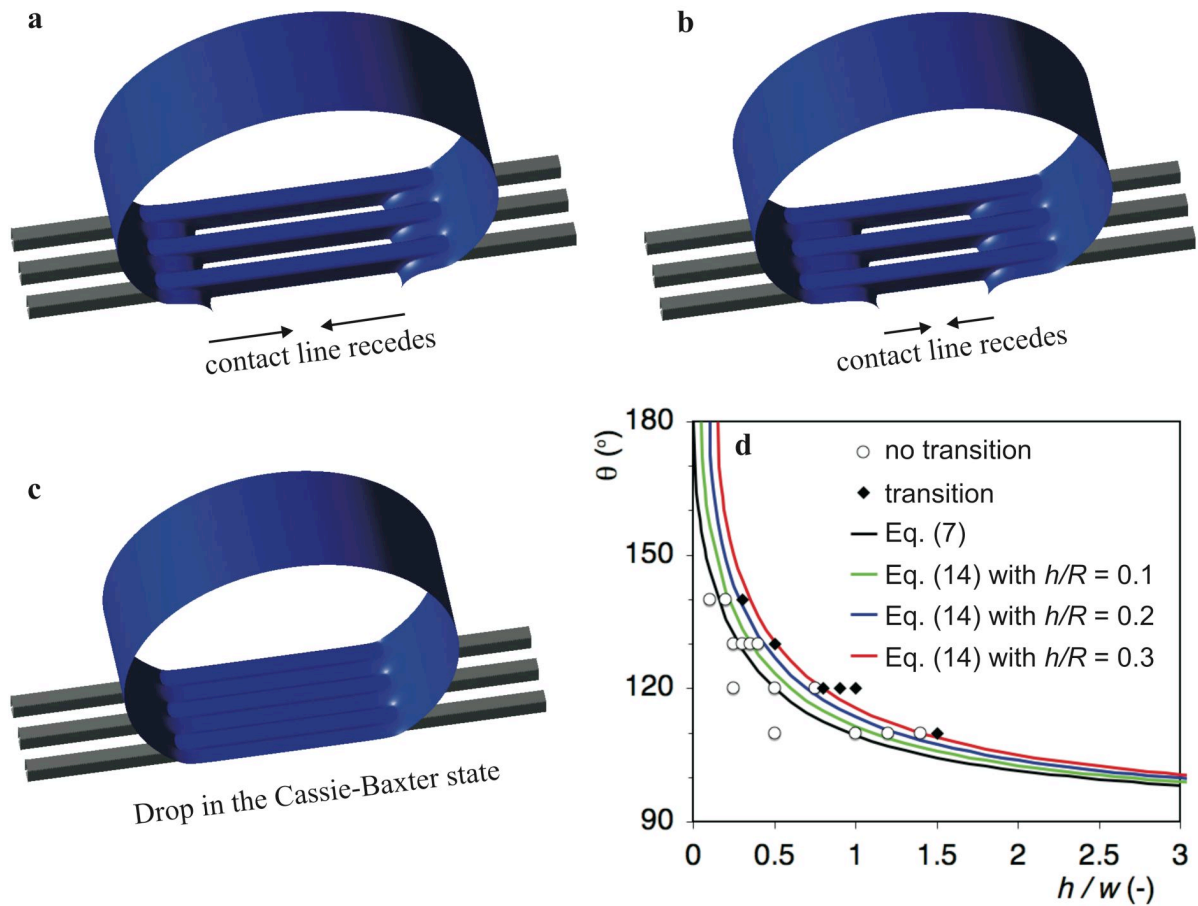

Figure 1: Lattice Boltzmann simulation results showing a lifting transition. The simulation results for cylindrical drops on parallel grooves: the drop dewets the grooves and undergoes a lifting transition. The shape of the cylindrical drop is shown at (a) $t = 10^5$, (b) $1.5 \times 10^5$, and (c) $2.5 \times 10^5$. (d) The lifting transition diagram as a function of groove aspect ratio and contact angle. The solid line represents the theoretical prediction of equations (7) and (14). The lattice Boltzmann simulations results are shown in diamonds, where the transition occurs, and in circles where it does not occur.

For the geometry shown in the Figure 2, the Cassie-Baxter contact angle is given by

$$\cos\theta_{CB} = \frac{\eta\cos\theta_e - 1}{\eta + 1}, \qquad (8)$$

where $\Phi = \eta/(\eta+1)$, $\eta = w_0/w$ and the Young's angle $\theta_e$ of the surface is defined as

$$\cos\theta_e = \frac{\gamma_{SG} - \gamma_{SL}}{\gamma_{LG}}. \qquad (9)$$

The $\gamma$'s correspond to the surface tensions and the subscripts S, L and G denotes the solid, liquid and gas respectively.

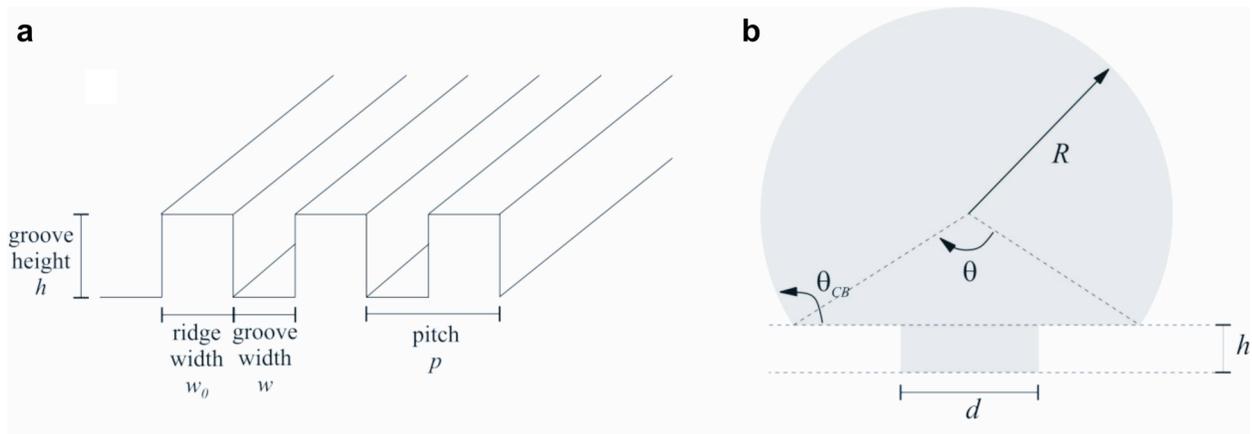

Figure 2: The schematic diagram of the geometry. (a) The surface pattern viewed along the corrugation. (b) The side view of the drop. The parameters of the model are defined in the figures.

Under the assumptions we have described above, the total volume is given by

$$V = w\left[\frac{1+\eta}{2}(2\pi - \theta + \sin\theta)R^2 + hd\right], \qquad (10)$$

and the total energy reads (we have subtracted the energy of the bare substrate):

$$E = (2\pi - \theta)(1 + \eta)wR\gamma_{LG} +$$
$$+ (2R\sin(\theta/2) - d)w[\gamma_{LG} + \eta(\gamma_{SL} - \gamma_{SG})]$$
$$+ 2hw\gamma_{LG}$$
$$+ d[(w(1+\eta) + 2h)(\gamma_{SL} - \gamma_{SG})]$$
(11)

The four terms in the above equation account for the liquid-gas interface of the cylindrical cap, the base of the drop in contact with the composite solid/gas substrate, the liquid-gas interface in the channel, and the liquid-solid interface in the channel respectively. The angle $\theta$ is related to the Cassie-Baxter angle via

$$\theta = 2(\pi - \theta_{CB}).$$
(12)

The Wenzel to Cassie-Baxter lifting transition occurs when $\partial_d E > 0$. Calculating the first derivative of $E$ with respect to $d$ and taking into account the volume constraint, we obtain

$$\partial_d E = w\gamma_{LG}[-h/R - 1 - (1 + 2h/w)\cos\theta_e].$$
(13)

The lifting transition requirement $\partial_d E > 0$ leads to the criterion:

$$\cos\theta_e < -\frac{1 + h/R}{1 + 2h/w}.$$
(14)

Compared to equation (7), the finite size effect of the drop volume adds an extra term $h/R$ in the numerator. In Figure 1d we plot equation (14) for several values of $h/R$. The match between simulation results and analytical theory is improved considerably; the $h/R$ term shifts the lifting transition to higher aspect ratio for a given intrinsic (Young's) contact angle. In other words, a smaller drop requires a more extreme geometry (higher aspect ratio) to enable the transition from the collapsed to the suspended state.

Qualitatively, our results provide a manufacturing guideline. For a lifting transition to occur there must be a dynamical pathway on which there is little contact line pinning (or contact angle hysteresis). When pinning or hysteresis is a dominant factor, e.g. in the direction perpendicular to the grooves or when the surface is patterned with rectangular posts, this type of dynamical pathway cannot be realized and the lifting transition cannot occur. Indeed, for the full three dimensional simulations of a lifting transition, we observe that the contact line *only* dewets in the parallel direction. Perpendicular to the grooves, the contact line is pinned. Equations (7) and (14) also dictate that the grooves aspect ratio must be high enough so that the suspended state is preferable.

In Supporting Movie 1, a lattice Boltzmann simulation of the lifting transition is shown for a full, three-dimensional drop of radius 110 lattice spacings. At the beginning of the movie, the drop is in the Wenzel state. The surface's intrinsic contact angle is 80 degrees and the aspect ratio of the corrugations is $h/w$ = 12/8 = 1.5. The intrinsic contact angle is then changed to 120 degrees and the drop undergoes a transition to the Cassie-Baxter state. Initially, the drop relaxes to a more spherical shape. Then, after approximately 1-2 s in the movie, the contact line on top of the barriers does not move considerably anymore. The liquid-vapor fronts inside the grooves, however, continue to recede. In other words, the channels are filled with vapor. This motion can be visualized by observing the lighter blue wetted parts of the channel. These regions decrease with time until finally the two liquid-vapor interfaces annihilate each other. The lifting transition is now completed.

*Experimental methods*

Photolithography was used to prepare corrugated surfaces with the desired aspect ratio as well as reference samples. Indium thin oxide (ITO) coated glass surfaces were coated with a thin layer of SU-8 negative photoresist (MicroChem Corp.), followed by a heating step (1 minute at 65ºC and 2 minutes at 95ºC) which is flood exposed with UV light to promote adhesion of the SU-8 surface texture. After crosslinking (1 minute at 65ºC and 2 minutes at 95ºC), a second SU-8 layer with a thickness of 18 μm was applied. After a heating step, the SU-8 photoresist was patterned (30 µm line pattern) via standard photolithography to give an aspect ratio $h/w$ of approximately 1.4. A 300-500nm Teflon coating (Dupont) is applied from solution on top of the structure to increase the hydrophobicity of the surface. The contact angle of the Teflon coated surface is 110°.

Scanning electron microscopy (XL 30 ESEM-FEG, Philips) was used to image the surface structures. Before the SEM analysis, a 15 nm gold layer (K575 XD Turbo Sputter Coater, Emitech Ltd.) is applied on the sample to improve electron conduction. The aspect ratio was determined using confocal microscopy, (Sensofar, PLµ2300) with a 50x objective.

To investigate the wetting behavior of water droplets on the surfaces, the contact angle was measured (OCA-30, DataPhysics Germany) upon reversibly applying an electric potential of up to 150V, by means of attaching the anode to the ITO layer and inserting the cathode (a small conductive pin) in the droplet from the top. The voltage was supplied by a Delta Electronica ES0300-0.45 power supply. Common tap water was used, having a moderate loading of dissolved ions. The anode clamp was attached to the ITO layer on the side of the sample. The cathode was attached to a thin needle and inserted into the water drop from the top.

To directly and unambiguously observe the lifting transition, optical microscopy was performed using a Leica DM6000M (5x objective with 2x magnification), equipped with a Leica DFC 420C camera in transmission mode. Images from the movie were exported as JPEG via the VirtualDub editing freeware (version 1.9.0). The stills were prepared as TIFF figures in Corel PhotoPaint X3, converting to 32bit CMYK and resizing the image to the appropriate resolution.

*Experimental results*

To investigate the wetting behavior of water droplets on the corrugated Teflon surface (Figure 3a), the contact angle is measured upon reversibly applying an electric potential of 150V DC. The electric voltage is increased linearly from 0 to 150V in 30 seconds, kept at this voltage for 20 seconds for the drops to equilibrate, after which the voltage is reduced back to 0V in 30 seconds. Figure 3b-d shows the contact angle as a function of time on respectively a non-corrugated and corrugated Teflon coated surface. The water droplet goes through the above described electrowetting cycle multiple times. None of the surfaces displays any visual signs of degradation due to e.g. pinhole conductive paths in the coatings. The measurements show that the electrowetting effect on a smooth surface is nicely reproducible and with a large range of contact angles from 110 to 60 degrees. This smooth and repeatable transition is also achieved on the corrugated surface. When the voltage is 0V, the drop is in the suspended state, in agreement with equation (7). In fact, the value of the contact angle agrees well with the Cassie-Baxter angle,[5]

$$\cos\theta_{CB} = \Phi\cos\theta_e - (1-\Phi) \qquad (15)$$

where $\Phi$ is the solid fraction of the surface, and $\theta_e$ is the Young's angle of the surface. Using $\Phi \sim 0.5$ and $\theta_e \sim 110$ degrees, we obtain $\theta_{CB} \sim 130$ degrees.

On the other hand, when the electrowetting potential is switched on, the surface behaves as a hydrophilic surface with an intrinsic contact angle of 60 degrees. The suspended state is no longer preferable and the liquid drop collapses. It should be noted that the drop shape in the collapsed state is very elongated along the direction of the corrugations, and the contact angles are considerably different parallel and perpendicular to the corrugations. The contact angles measured perpendicular to the corrugations (i.e. viewed along the corrugations) are shown in Figure 3c as a function of time. The contact angle varies from 130 to 90 degrees. The contact angle measurements parallel to the corrugations are shown in Figure 3d for one electrowetting cycle. In the Wenzel state, contact angle measurements yield distinct results parallel (~50 degrees) and perpendicular (~ 90 degrees) to the grooves. This is due to pinning of the contact line on the ridges, which prevents the drop from spreading out perpendicular to the channels.[14] In Figure 4, this difference in elongation is shown from images captured by the OCA-30 contact angle setup during experiments.

Another important aspect of drops on surfaces is their sliding angle, i.e. the tilt angle at which a drop spontaneously starts moving due to gravity. While a fully controlled sliding angle experiment was difficult to perform since the cathode needle had to be kept properly inserted in the drop, our preliminary results confirm our expectations. Drops in the Wenzel state are stuck on the surface even when the tilt angle is increased to up to 90 degrees. In contrast, drops in the Cassie-Baxter state exhibit motion for a sliding angle of less than 10 degrees. We also observe no significant difference between the parallel and the perpendicular sliding angle for drops in the Cassie-Baxter state. This line of investigation is being further pursued and the results will be presented elsewhere.

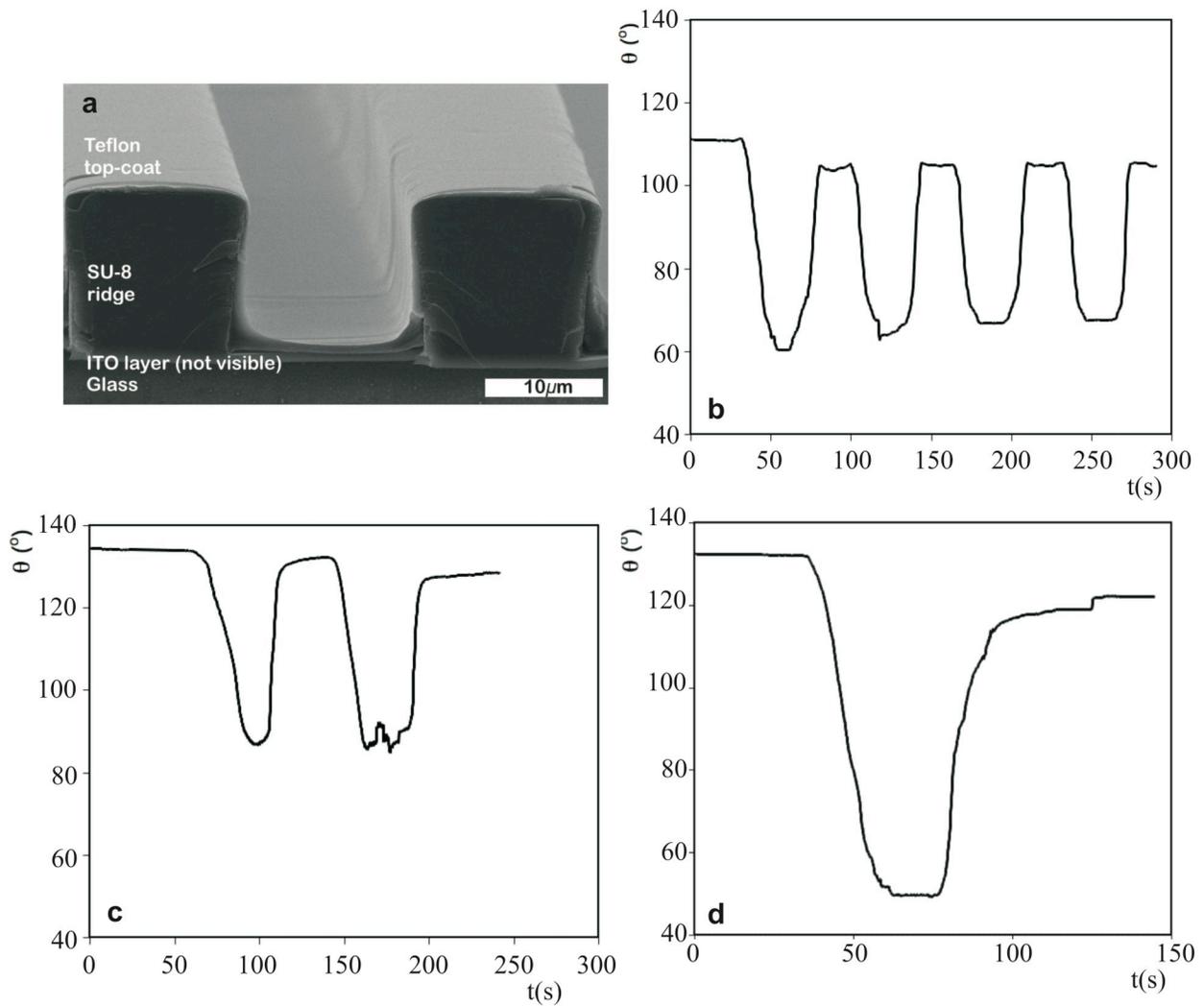

Figure 3: SEM cross-section of a corrugated surface used in the electrowetting experiments (a). Contact angle measurements on a non-patterned (b) surface show that the contact angle varies between 110 and 60°, whereas on a patterned (c-d) surface the contact angle can be reversibly switched between 130 and 90°(perpendicular)/50°(parallel) during electrowetting cycles. Contact angle measurements were done by automated contact angle fitting, employing an ellipsoidal drop profile. Using this method, the typical fitting error is ± 3°. The measurements in (c) and (d) are taken perpendicular and parallel to the grooves. Slight hysteresis does occur due to surface defects.

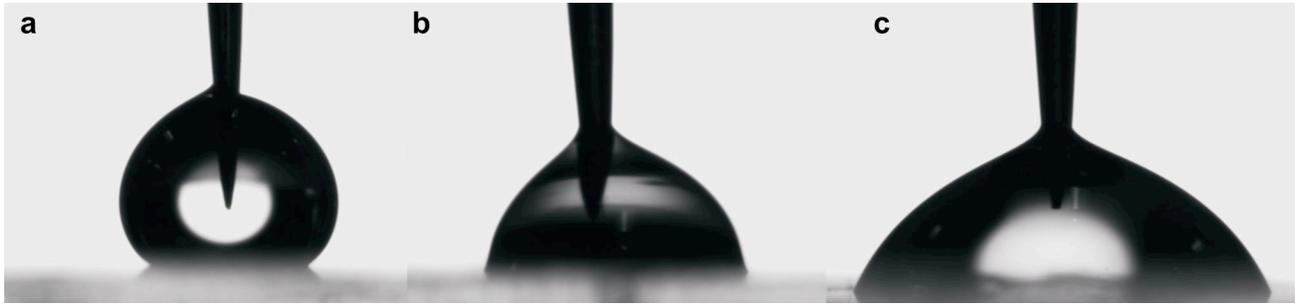

*Figure 4: Side view of water drops during electrowetting experiment on a corrugated surface. The needle is seen inserted from the top. (a) The Cassie-Baxter (suspended) drop at 0V before actuation, with a base radius of approximately 1mm. The Wenzel (collapsed) drop at 150V during electrowetting, viewed (b) perpendicular and (c) parallel to the grooves. The elongation of the drop parallel to the channels is clearly visible from the difference in base length and contact angle between (b) and (c).*

As further evidence that the lifting transition has occurred, optical microscopy was performed on the drop while the drop underwent subsequent collapse and uplifting transitions. In Supporting Movie 2, we present the behavior of a drop on a corrugated Teflon surface upon the application of an electrowetting cycle. The movie is sped up 10x. At the beginning (0-3 s) of the movie, we apply an electrowetting potential of 150 V. The water drop undergoes a collapse transition. This can be inferred from the outward movement of the liquid-vapor interface fronts in the channels. These fronts are black as they reflect light away from the optical path. Furthermore, the drop elongates along the direction of the channels. After maintaining the voltage at 150V briefly, the voltage is reduced to 0V (4-10 s). Like in Supporting Movie 1, the drop relaxes to a more spherical shape (higher contact angle state). We reposition the sample stage to keep the drop in the frame. The cathode needle is visible on the upper left corner. Between 11-30 s, we can see the receding motion of the contact line (liquid-vapor fronts) in the channel. Contrary to the simulations, these fronts do not appear symmetrically from both ends of the channel, nor do they proceed at equal speeds. This is due to pinning of the interfaces in the channels, which is caused by defects and surface roughness in the channels. Nonetheless, one can clearly see in a number of channels that the fronts from each side of the drop collide and annihilate one another. At the end of the movie, all channels are filled with vapor and the lifting transition is complete. In real time, the collapse and lifting transitions take place in the order tens of seconds and minutes respectively. The most relevant frames of the experimental movie are shown in Figure 5.

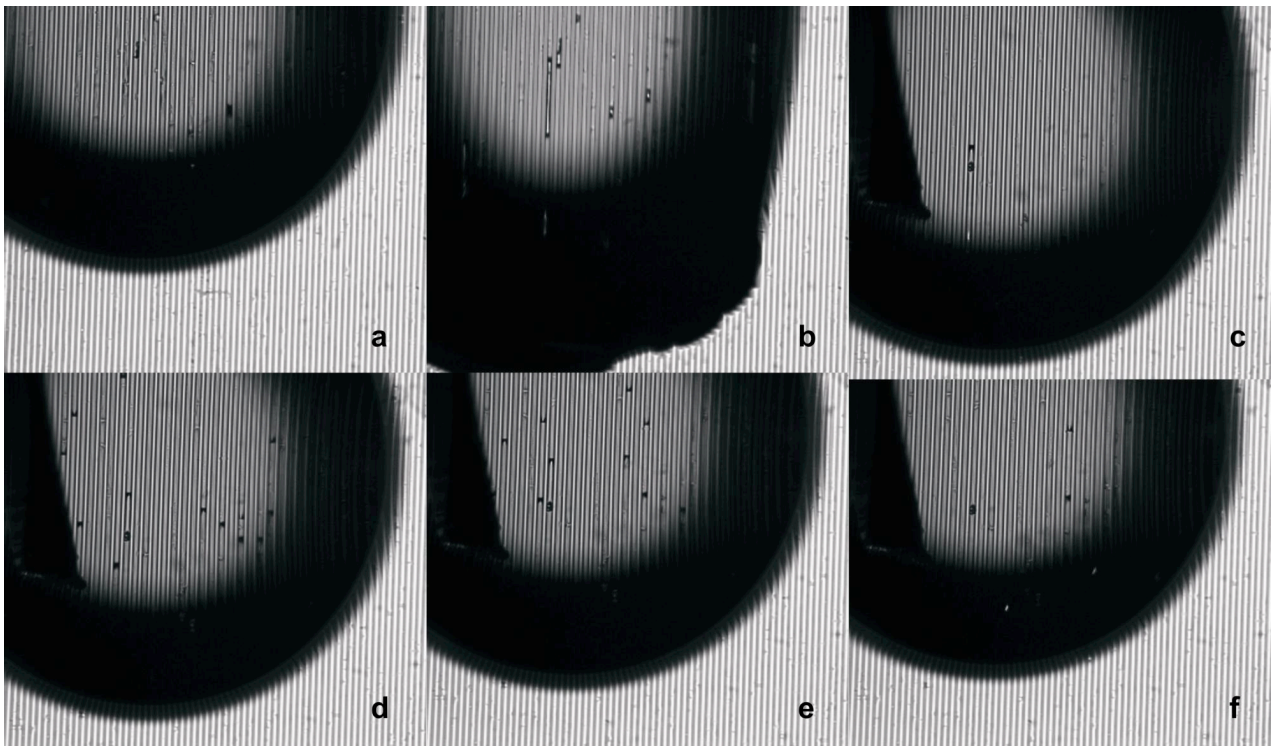

*Figure 5: Optical microscopy images of the subsequent collapse and lifting transitions of a water drop. (a-f) The collapse and lifting transition can be directly observed by monitoring the air/liquid interfaces underneath a water drop during electrowetting. The images chronologically follow an electrowetting cycle. (a) The initial suspended drop in rest. (b) Upon applying 150V, the collapse transition occurs and the drop spreads out, mostly along the direction of the channel. (c) By reducing the voltage to 0V, the drop retracts to a less spread out collapsed (Wenzel) drop. (d) Then the lifting transition starts, as indicated by darkened air/liquid interfaces in a number of the channels, moving in from the side of the drop along the channel. (e) The drop is now in an intermediate state, where some of the air/liquid interfaces have annihilated, as observed in the simulations, and others continue to dewet the channel. (f) Finally, all channels are filled up with air, completing the lifting transition and leaving the drop in the suspended (Cassie-Baxter) state again.*

## *Conclusions*

We have demonstrated that repeatable and reversible transitions between the Cassie-Baxter and Wenzel states can be induced by using a combination of a corrugated surface and electrowetting. The directionality of the surface patterning is key here as it allows the contact line to dewet the grooves with little or no energy barrier in the parallel direction, while it remains pinned in the perpendicular direction. We use lattice Boltzmann simulations to guide us in determining the required conditions and we realize the transition experimentally on a Teflon coated surface patterned with parallel grooves. The transitions are verified by contact angle measurements and by direct observation using an optical microscope. Since both surface patterning and electrowetting are already commonly used, we believe the design paradigm we present here is highly practical and it offers a clear opportunity to manipulate fluids in potential areas of applications, such as in the design of smart materials in microfluidics.

*Supporting Information Available*

We provide two movies which show the collapse and lifting transitions. The first movie is obtained from a three dimensional lattice Boltzmann simulation, while the second movie is an optical microscopy movie made during an actual collapse and uplifting electrowetting cycle.

*Acknowledgements*

The researchers would like to thank Hanneke Thijs for help with the contact angle setup and Prof. J.M. Yeomans for helpful discussions.

## SUPPORTING INFORMATION

*Collapse and lifting transition movies*

In order to clearly demonstrate the dynamical pathway of the lifting transition, we have prepared two Supporting Movies.

Supporting Movie 1 was prepared as follows. The raw data from the lattice Boltzmann simulation was visualized using (commercial) Matlab R2008b software. The snapshots were saved as tiff files, converted to png files using the standard Linux "convert" command, and combined using (open source) MEncoder software to create an MPEG-4 movie. We then cropped the movie with (freeware) VirtualDub 1.9.0 to remove excess white spaces. Finally the movie was recompressed with (open source) MEncoder.

Supporting Movie 2 was originally obtained from the Leica microscope as an uncompressed avi file at 100 frames per second. With the aid of (freeware) VirtualDub 1.9.0, we converted the movie to 8-bit grayscale, cropped the margins around the drop, and reduced the resolution of the movie with a 2:1 reduction filter. Three subsequent encoding steps were performed to reduce the file size. First, an Indeo Video 5.10 codec was employed in VirtualDub with a 1.0 temporal quality factor and a 300kbyte/second target data rate. Then, using Ulead Videostudio 9.0 the movie was sped up 10x and recompressed with a DivX 6.8.4 codec. Finally, it was recompressed using MEncoder to give an MPEG-4 movie. We made sure that no alteration in the contrast or intensity occurred and that no visual artifacts were introduced. The width of the supporting movie 2 corresponds to 2.1mm.